\documentclass[11pt,twoside]{article}

\usepackage{asp2006}
\usepackage{epsf}
\usepackage{psfig}
\usepackage{lscape}

\begin{document}
\title{Jet spectra in FRI radio galaxies: implications for particle
acceleration} %%% Fill in title 
\author{R.A. Laing} %%% Fill in author names
\affil{ESO, Karl-Schwarzschild-Stra\ss e 2, 85748 Garching-bei-M\"{u}nchen,
Germany}
%%% Fill in author affiliations
\author{A.H. Bridle, W.D. Cotton}
\affil{NRAO, 520 Edgemont Road, Charlottesville, VA
22903-2475, U.S.A.}    %%% Fill in author affiliations
\author{D.M. Worrall, M. Birkinshaw}
\affil{Department of Physics, University of Bristol, Tyndall Avenue,
Bristol BS8~1TL, U.K.}

\begin{abstract} %%% Abstract to run on from here.
We describe very accurate imaging of radio spectral index for the inner jets in
three FRI radio galaxies. Where the jets first
brighten, there is a remarkably small dispersion around a spectral index of
0.62. This is also the region where bright X-ray emission is detected. Further
from the nucleus, the spectral index flattens slightly to 0.50 - 0.55 and X-ray
emission, although still detectable, is fainter relative to the radio. The
brightest X-ray emission from the jets is therefore not associated with the
flattest radio spectra, but rather with some particle-acceleration process whose
characteristic energy index is 2.24.The change in spectral index occurs roughly
where our relativistic jet models require rapid deceleration.  Flatter-spectrum
edges can be seen where the jets are isolated from significant surrounding diffuse
emission and we suggest that these are associated with shear.
\end{abstract}

%%% MAIN BODY OF TEXT GOES HERE. CONSULT "INSTRUCTIONS FOR AUTHORS USING
%%% LATEX2E MARKUP", SECTIONS 2.3-2.6 FOR HELP WITH EQUATIONS, FIGURES,
%%% AND TABLES.

%\section{}   %%% Top level section head (remove "%" symbol)
%\subsection{}   %%% Second level section head (remove "%" symbol)
%\subsubsection{}   %%% Lowest level section head (remove "%" symbol)
%\section*{}    %%% Unnumbered top level section head (remove "%" symbol)
%\subsection*{}   %%% Unnumbered second level section head (remove "%" symbol)

\section{Radio spectra: 3C\,31}
\label{radio}

The detection of X-ray synchrotron emission on kiloparsec scales in the bases of
low-luminosity (FR\,I) radio jets (e.g.\ \citealt{Hard02,Hard05,ngc315xray})
requires distributed particle acceleration, consistent with the failure of
adiabatic models to reproduce the observed brightness profiles at radio
wavelengths \citep{LB04}. The acceleration mechanism is not understood. Here, we
summarize the results of recent work on this problem, using accurate radio
spectral mapping (this section) and detailed radio -- X-ray comparisons
(Section~\ref{xray}).  

\begin{figure}
\plotone{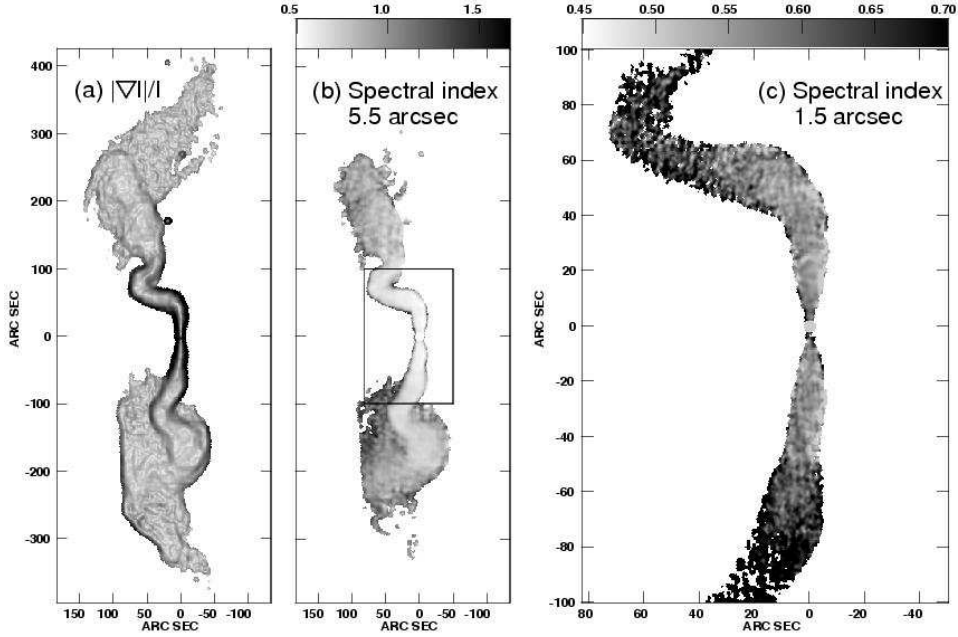}
\caption{Radio images of 3C\,31. (a) Sobel-filtered, mean L-band image
  (normalized by total intensity) at a resolution of 5.5\,arcsec.  (b) and (c)
  Spectral index, $\alpha$ from weighted least-squares, power-law fits to the
  total intensity.  (b) 5-frequency fit between 1365 and 4985\,MHz at a
  resolution of 5.5\,arcsec FWHM.  (c) 6-frequency fit between 1365 and
  8440\,MHz at a resolution of 1.5\,arcsec FWHM for the inset area in panel
  (b).\label{fig:3c31alpha}}
\end{figure}

In the course of our jet-modelling programme \citep[and references
therein]{L07}, we have accumulated high-resolution, multi-frequency radio images
of the bases of FR\,I jets and have used these to derive accurate spectral-index
distributions.  We have studied NGC\,315 \citep{ngc315ls} and 3C\,296
\citep{3c296}; a third example, the nearby radio galaxy 3C\,31 (z = 0.0169;
Laing et al., in preparation), is shown in Fig.~\ref{fig:3c31alpha}. There is no
evidence for any significant deviation from power-law spectra in 3C\,31 within
70\,arcsec of the nucleus.  Out to $\approx$7\,arcsec in both jets, the spectral
index at 1.5-arcsec resolution is slightly steeper ($\langle \alpha \rangle =
0.62$)\footnote{$S(\nu) \propto \nu^{-\alpha}$} than the average for the inner
jets. From 7 -- 50\,arcsec in both jets, the mean spectral index is in the range
0.55 -- 0.57. Further from the nucleus, there is a gradual spectral steepening.
There are also slight, but significant variations in spectral index across both
jets within $\approx$30\,arcsec of the nucleus in the sense that their West
edges tend to have flatter spectra ($\langle \alpha \rangle = $ 0.52 -- 0.54;
Fig.~\ref{fig:3c31alpha}c).  There is a clear spectral separation between the
South jet and the surrounding emission, matching the separation of these regions
defined by the sharpest brightness gradients (Figs~\ref{fig:3c31alpha}a and b).
This is particularly clear where the jet first enters the diffuse emission. The
spectral identity of the jet is evidently maintained even after it bends
abruptly about 2\,arcmin South of the nucleus and remains until it terminates in
a region of high brightness gradient.  The outer South jet in 3C\,31 is
therefore a clear example of the type of spectral structure noted in other FR\,I
sources \citep{KSR97,3c296} wherein a flatter-spectrum `jet' with a distinct
spectral identity is superposed on a `sheath' of steeper-spectrum emission.

\section{Radio -- X-ray comparison: NGC\,315}
\label{xray}

\begin{figure}
\plotone{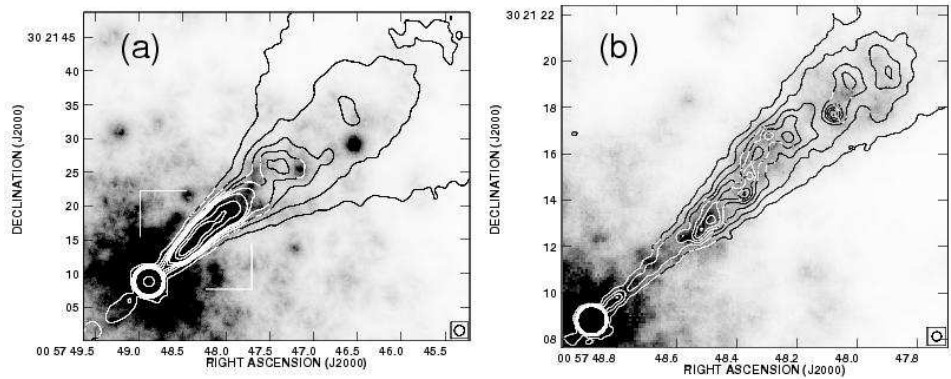}
\caption{Overlays of VLA 5\,GHz images (contours) on adaptively smoothed {\em
    Chandra} 0.8 -- 5\,keV data (grey-scale) for NGC\,315 \citep{ngc315xray}. (a) Radio
    resolution 1.5\,arcsec. The area covered by panel (b) is shown by the white
    box. (b) radio resolution 0.4 arcsec.\label{fig:ngc315xray}}
\end{figure}

\begin{figure}
\plotone{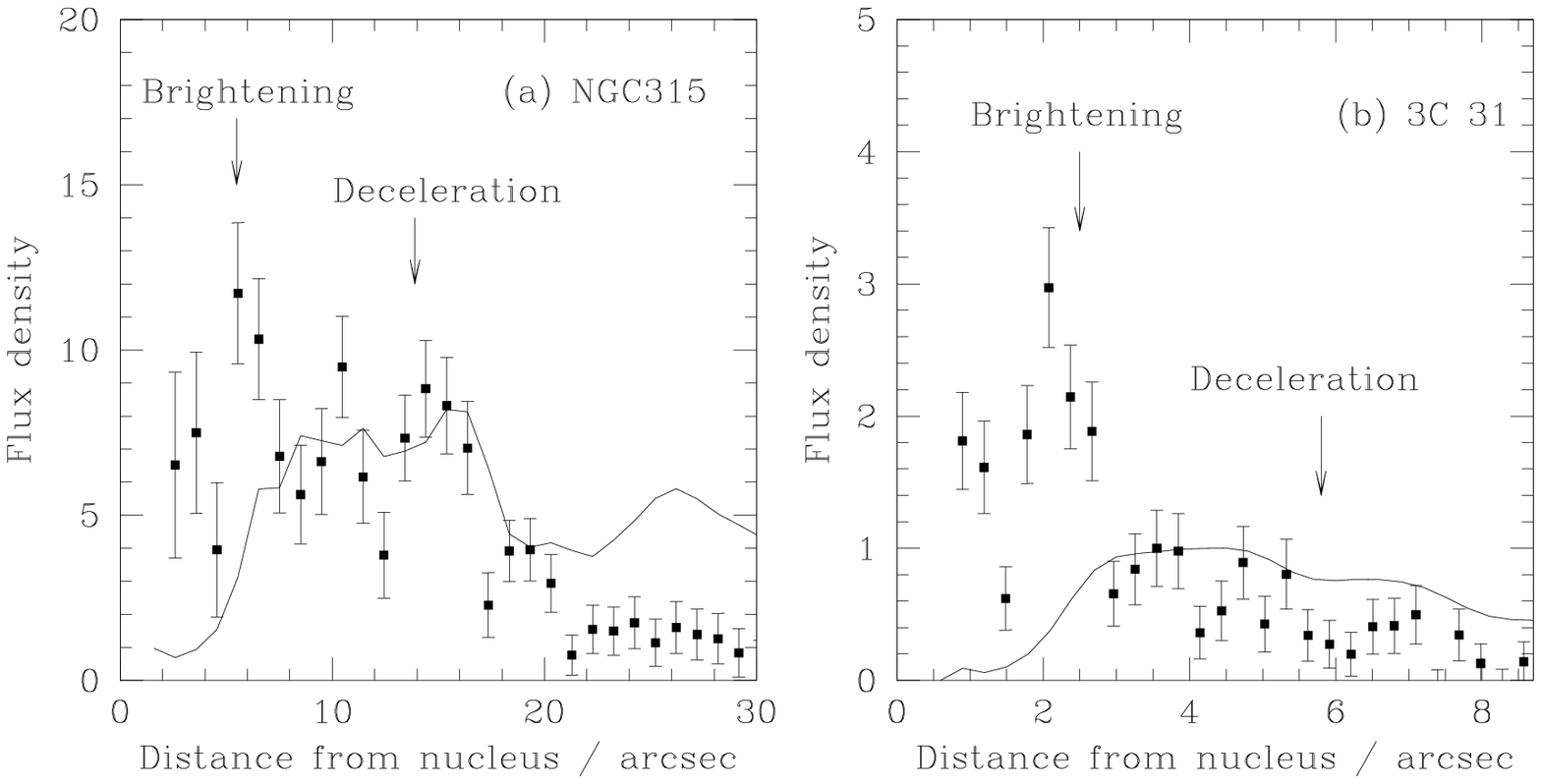}
\caption{Profiles of radio (curve) and X-ray (points) flux density along the
  brighter jets of NGC\,315 \citep{ngc315xray} and 3C\,31 \citep{LB04}. The locations of
  the brightening points (where the rest-frame radio emission increases rapidly)
  and the start of rapid deceleration, as modelled by \citet{CLBC} and \citet{LB02}
  are also indicated.\label{fig:xprofiles}}
\end{figure}

Comparison of deep, high-resolution radio and X-ray images also provides clues
to particle-acceleration processes.  Fig.~\ref{fig:ngc315xray} shows overlays of
5-GHz radio emission at resolutions of 1.5 and 0.4\,arcsec FWHM on deep {\em
Chandra} X-ray images for the main jet in NGC\,315 (z = 0.01648;
\citealt{ngc315xray}).  The radio and X-ray spectral indices integrated over the
inner jet are 0.61 and 1.2, respectively, consistent with synchrotron emission
from a single population of relativistic electrons in both wavebands.  The radio
jet is relatively faint and unresolved in width out to about 4\,arcsec from the
core, after which it brightens abruptly. The most prominent feature of the radio
brightness distribution between 4 and 12\,arcsec (Fig.~\ref{fig:ngc315xray}b) is
an oscillatory filament, whose nature is discussed in detail by
\citet{ngc315xray}.  The first X-ray enhancement which can unambiguously
associated with the jet is 3.6\,arcsec from the nucleus, where the radio
emission is still faint. From 5 -- 7.5\,arcsec, there is particularly good
morphological correspondence between radio and X-ray images, with diffuse
emission over the full width of the jet as well as localized X-ray peaks
following the radio ridge-line. At larger distances, the general correspondence
is still reasonable, but there are differences of detail. The X-ray emission can
be traced out to $\approx$30\,arcsec (Fig.~\ref{fig:ngc315xray}a), but becomes
weaker with respect to the radio at $\approx$20\,arcsec. This is most clearly
illustrated by the profiles of X-ray and radio emission along the jet at matched
resolutions in in Fig.~\ref{fig:xprofiles}; an equivalent plot for 3C\,31
\citep{LB04} is shown for comparison.

\section{Discussion}

These observations contribute to the developing picture of spectral variations
in FR\,I sources. Bright X-ray emission is detected close to the nucleus, in the
faint, well-collimated jet bases that precede the sudden radio brightening
(e.g.\ Fig.~\ref{fig:xprofiles}). There is approximate morphological
correspondence between features in the radio and X-ray brightness distributions
after the former brightens, although there are differences on small scales
(e.g.\ Fig.~\ref{fig:ngc315xray}b). In contrast, there are no systematic
transverse variations in the X-ray/radio ratio within $\approx$30\,arcsec of the
nucleus in NGC\,315 (the best resolved case; \citealt{ngc315xray}). Particle
acceleration appears to be distributed throughout the jet volume, rather than
being exclusively associated with discrete knots or with the boundary.  The
ratio of X-ray to radio emission decreases where our kinematic models show that
the jets start to decelerate from speeds of 0.8 -- 0.9$c$
(Fig.~\ref{fig:xprofiles}; \citealt{LB02,Hard02,CLBC,ngc315xray}).

Where the jets first brighten and before they decelerate, there is a remarkably
small dispersion around a radio spectral index of $\alpha = 0.62$ in the three
sources we have studied in detail, as well as 3C\,66B (Fig.~\ref{fig:3c31alpha};
\citealt*{HBW,ngc315ls,3c296}).  The average is dominated by emission immediately
after the point at which the jets first brighten. This is also the region from
which X-ray emission is detected from the main jets in all four sources
(Fig.~\ref{fig:xprofiles}; \citealt{HBW,Hard05}).  The spectral index of the fainter
emission close to the nucleus in 3C\,449 \citep{KSR97}, PKS1333$-$33 \citep{KBE}
and 3C\,66B \citep{HBW} appears to be slightly steeper than $\alpha = 0.62$
although the uncertainties are larger. Further from the nucleus, the spectra
flatten slightly to $\alpha = $ 0.50 -- 0.55, contrary to any naive expectation
from models in which electrons are accelerated at the brightening point and
suffer synchrotron losses as they propagate. X-ray emission is still detected
from these regions, but at a lower level relative to the radio
(Fig.~\ref{fig:xprofiles}). The brightest X-ray emission from the jets is
therefore not associated with the flattest radio spectra, but rather with some
particle acceleration process whose characteristic energy index is $2\alpha + 1
= 2.24$. A related result is that an asymptotic low-frequency spectral index of
0.55 is common in FR\,I jets over larger areas than we consider here
\citep{Young}. Flatter-spectrum edges can be seen where the jets are isolated
from significant surrounding diffuse emission, most clearly in NGC\,315
\citep{ngc315ls}. Our kinematic models \citep{LB02,CLBC,3c296} show that all of
the jets have substantial transverse velocity gradients and it is plausible that
the process that produces the flatter spectrum is associated with high shear
\citep{SO}.  In 3C\,31, the flatter-spectrum regions (Fig.~\ref{fig:3c31alpha}c)
occur predominantly on the outer edges of bends, perhaps consistent with this
idea.

As well as a smooth steepening of the jet spectrum at larger distances from the
nucleus, as would be expected from synchrotron and adiabatic losses affecting a
homogeneous electron population, multiple spectral components are observed
(Fig.~\ref{fig:3c31alpha}b). Jets appear to retain their identities even after
entering regions of diffuse emission and are clearly identifiable by their
flatter spectra. They are usually also separated from the surrounding emission
by sharp brightness gradients (Fig.~\ref{fig:3c31alpha}a).  This spine/sheath
separation is observed in FR\,I sources with bridges of emission extending back
towards the nucleus (e.g.\ 3C\,296; \citealt{3c296}) as well as tailed sources
like 3C\,31 (Fig.~\ref{fig:3c31alpha}). Although there is an overall trend for
the spectrum of the diffuse emission to steepen towards the nucleus in bridges
and away from it in tails \citep{Parma99}, the variations in individual objects
are complex.  The termination regions of jets in tailed FR\,I sources are
perhaps best regarded as bubbles which are continually fed with fresh
relativistic plasma by the jets and which in turn leak material into the tails.
Their spectral steepening would then be governed by a combination of continuous
injection, adiabatic, synchrotron and inverse Compton energy losses and escape.

\acknowledgements %%% Text of acknowledgements runs on after this command.
The National Radio Astronomy Observatory is a
facility of the National Science Foundation operated under cooperative agreement
by Associated Universities, Inc.

%%% THE BIBLIOGRAPHY
%%%
%%% CONSULT SECTION 3 OF "INSTRUCTIONS FOR AUTHORS" FOR HOW TO USE NATBIB.
%%% AUTHORS ARE ENCOURAGED TO USE EITHER THE "THEBIBLIOGRAPY" ENVIRONMENT
%%% BY UNCOMMENTING (DELETING THE "%" SYMBOL) THE COMMANDS BELOW, OR BY
%%% USING THE BIBTEX ENVIRONMENT. TO FIND OUT WHICH IS APPLICABLE TO YOUR
%%% CONTRIBUTION, CONSULT THE VOLUME EDITORS FOR YOUR PROCEEDINGS.
%%%

\end{document}